\documentclass[conference]{IEEEtran}
\IEEEoverridecommandlockouts
\usepackage{tabularx,booktabs}
\usepackage{multirow}

\usepackage{cite}
\usepackage{amsmath,amssymb,amsfonts}
\usepackage{algorithmic}
\usepackage{graphicx}
\usepackage{textcomp}
\usepackage{xcolor}
\def\BibTeX{{\rm B\kern-.05em{\sc i\kern-.025em b}\kern-.08em
    T\kern-.1667em\lower.7ex\hbox{E}\kern-.125emX}}
\begin{document}

\title{Exploring In-Context Learning Capabilities of ChatGPT for Pathological Speech Detection 
\thanks{This work was supported by the Swiss National Science Foundation project CRSII5\_202228 on ``Characterisation of motor speech disorders and processes''.}
}

\author{\IEEEauthorblockN{Mahdi Amiri$^{1 ,2}$, Hatef Otroshi Shahreza$^{1}$, Ina Kodrasi$^{1}$}
\IEEEauthorblockA{
\textit{$^1$Idiap Research Institute, Switzerland} \\
\textit{$^2$École Polytechnique Fédérale de Lausanne (EPFL), Switzerland}\\}

{\tt \{mahdi.amiri,hatef.otroshi,ina.kodrasi\}@idiap.ch}
% \and
% \IEEEauthorblockN{Hatef Otroshi Shahreza$^{*}$}
% \IEEEauthorblockA{\textit{dept. name of organization (of Aff.)} \\
% \textit{name of organization (of Aff.)}\\
% City, Country \\
% email address or ORCID}
% \and
% \IEEEauthorblockN{Sébastien Marcel}
% \IEEEauthorblockA{\textit{dept. name of organization (of Aff.)} \\
% \textit{name of organization (of Aff.)}\\
% City, Country \\
% email address or ORCID}
% \and
% \IEEEauthorblockN{Ina Kodrasi}
% \IEEEauthorblockA{\textit{Signal Processing for Communication Group} \\
% \textit{Idiap Research Institute}\\
% Martigny, Switzerland \\
% ina.kodrasi@idiap.ch}

}

\maketitle

\begin{abstract}

Automatic pathological speech detection approaches have shown promising results, gaining attention as potential diagnostic tools alongside costly traditional methods. While these approaches can achieve high accuracy, their lack of interpretability limits their applicability in clinical practice. 
%Recently, with the rise of Large Language Models (LLMs), many researchers have started leveraging these models as the backbone of their proposed methods. 
In this paper, we investigate the use of multimodal Large Language Models (LLMs), specifically ChatGPT-4o, for automatic pathological speech detection in a few-shot in-context learning setting. 
Experimental results show that this approach not only delivers promising performance but also provides explanations for its decisions, enhancing model interpretability. To further understand its effectiveness, we conduct an ablation study to analyze the impact of different factors, such as input type and system prompts, on the final results.
Our findings highlight the potential of multimodal LLMs for further exploration and advancement in automatic pathological speech detection.
% This paper investigates the use of multimodal large language models (LLMs), specifically ChatGPT-4o, for automatic pathological speech detection in a few-shot in-context learning setting. Pathological speech, resulting from neurological disorders such as Cerebral Palsy, Amyotrophic Lateral Sclerosis, and Parkinson's disease, often leads to speech impairments like dysarthria and apraxia, significantly affecting communication. Traditional diagnostic methods, which rely on auditory-perceptual assessments, are time-consuming and resource-intensive. In recent years, Deep Learning (DL) approaches have shown promise in automating this process, but their lack of interpretability limits their applicability in critical healthcare settings. In contrast, multimodal LLMs offer improved transparency and interpretability, enabling better understanding of model decision-making. Our experiments demonstrate that ChatGPT-4o, leveraging Short-Time Fourier Transform (STFT) representations, performs comparably to state-of-the-art methods in pathological speech detection, even with limited labeled data. These findings highlight the potential of multimodal LLMs for advancing automatic speech assessment systems in clinical settings.
\end{abstract}


\section{Introduction}
Pathological speech can result from neurological damage caused by conditions such as Cerebral Palsy, Amyotrophic Lateral Sclerosis, or Parkinson’s disease. These disorders often lead to speech impairments such as dysarthria and apraxia of speech, which can significantly affect communication \cite{duffy2012motor, ziegler2012apraxia}. Traditionally, speech and language pathologists conduct auditory-perceptual assessments to diagnose these conditions in a clinical setting, which is both costly and time-consuming. 
To reduce this burden on healthcare systems, researchers are actively developing automated methods for detecting pathological speech. Earlier approaches combined handcrafted acoustic features with traditional machine learning techniques~\cite{handcrafted_1, handcrafted_2, handcrafted_3}. With the remarkable success of deep learning (DL) in various fields~\cite{transformer_ref, speech_ref}, efforts have increasingly shifted toward using DL-based approaches for automatic pathological speech detection~\cite{baseline, deep_learning_based, mel_ref_vae, mfcc_ref, schu2023using, lstm_patho_ref, amiri2024test, amiri_suppress}.

\begin{figure*}[t]
  \centering
  \includegraphics[width=\textwidth]{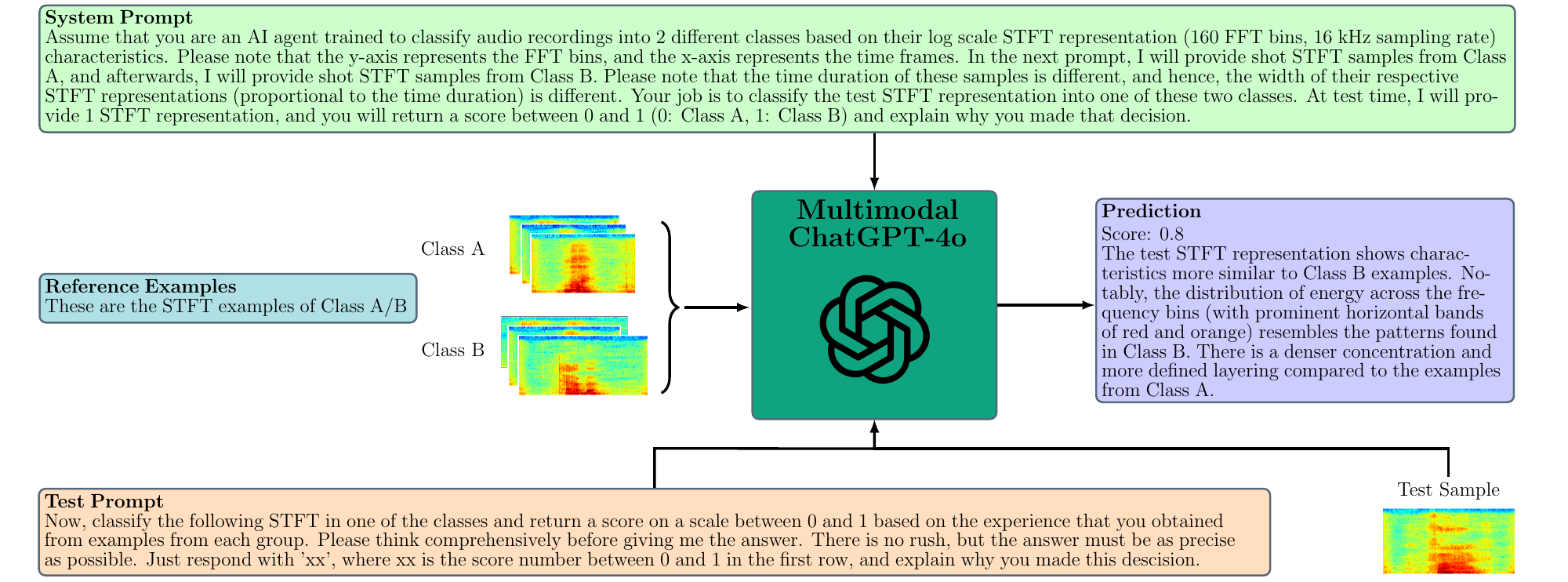}
  \caption{Schematic illustration of the proposed method. We set a system prompt that describes the classification task, input representation, and the number of reference samples per class. Then, several samples from each class are provided to the model, which is asked to classify the test sample based on them. In response, the model returns a classification score and explains the reasoning behind its decision.}
  \label{fig:measures}
\end{figure*}

Despite the remarkable performance of DL approaches across various domains, their interpretability remains a significant challenge~\cite{schu2023using, interpret}. Such approaches often function as black boxes, making it difficult to understand how they derive their decisions. This lack of transparency is particularly concerning in critical areas such as healthcare, where understanding the decision-making process is essential for trust and reliability.

With the rise of multimodal large language models (LLMs), new research directions are emerging that extend beyond traditional DL approaches. While these models were originally designed for natural language processing, they have been developed and extended for other domains~\cite{latif2023sparkslargeaudiomodels,awais2025foundation,fmbiometrics2025survey}. One of the key advantages of multimodal LLMs is their ability to explain the reasoning behind their decisions, providing interpretability in the process. These models can describe the factors and context that influence their predictions. 

Among these multimodal LLMs, ChatGPT-4o~\cite{gpt4} stands out as one of the most advanced models, demonstrating exceptional capabilities in understanding and processing different modalities (such as text and vision) among different applications~\cite{hassanpour2024chatgpt,koubaa2023exploring,he2023can,komaty2025exploring}. Moreover, the performance of these models can be improved on downstream tasks through few-shot in-context learning~\cite{brown2020language}. In few-shot in context learning, the weights of the model remains unchanged, but the model is prompted with a few examples before being asked about the test query.

% {\textcolor{red}{
% Given the promising in-context learning capabilities of ChatGPT, in this paper we investigate the performance of the multimodal ChatGPT-4o for pathological speech detection in a few-shot in-context learning scenario.  Given the advances in vision-based LLMs, this study focuses on analyzing the capabilities of ChatGPT-4o in processing STFT magnitude spectrogram representations for automatic pathological speech detection, which is the central objective of this paper. Additionally, we investigate the performance of our proposed method with raw audio input as part of an ablation study. 
% }}

Given the promising in-context learning capabilities of ChatGPT, in this paper, we investigate the performance of the multimodal ChatGPT-4o for pathological speech detection in a few-shot in-context learning scenario. Ideally, one should directly analyze raw speech inputs in the context of pathological speech detection; however, GPT-4o does not support direct audio input. While GPT-4o-audio-preview supports direct audio input, its overall capabilities relative to GPT-4o remain unclear, as it is a preview model. Therefore, we focus on evaluating GPT-4o’s ability to process short-time Fourier transform (STFT) magnitude spectrogram representations for pathological speech detection, making it the central objective of our paper. Nevertheless, to ensure completeness, we also provide an ablation study assessing the performance when using raw speech input and the GPT-4o-audio-preview model.

Experiments on the Noise Reduced UA-Speech database~\cite{ua-denoise}, which includes control and dysarthric speech from Cerebral Palsy patients, show promising results. Specifically, ChatGPT-4o achieves competitive performance compared to a state-of-the-art (SOTA) pathological speech detection approach that also operates on magnitude STFT spectrogram inputs~\cite{baseline, amiri2024adversarial}, despite having access to significantly less labeled training data.
Notably, while the SOTA model is trained from scratch with more data \cite{baseline, amiri2024adversarial}, it lacks the broad knowledge and pretraining of ChatGPT-4o.
Additionally, ChatGPT-4o provides explanations for its decisions, enhancing model interpretability, which is crucial for healthcare applications. These findings highlight the potential of multimodal LLMs for further exploration and advancement in automatic pathological speech detection. To the best of our knowledge, this work is the first to employ multimodal LLMs for pathological speech detection.
The key contributions of this paper are:
\begin{itemize}
    \item We propose a method for utilizing multimodal LLMs in automatic pathological speech detection.   
    \item We evaluate our proposed method against a SOTA baseline, demonstrating that it offers promising and competitive performance, while providing explanations.
    %\item We investigate the effect of different representations (STFT magnitude spectrogram and raw speech) on the final results. 
    %Our results demonstrate that using STFT representation leads to better results.
    \item We conduct an ablation study to further analyze the effect of different factors on the performance of the proposed method.
\end{itemize}

\section{Related Works}

\subsection{DL-based Automatic Pathological Speech Detection}
Traditionally, DL-based automatic pathological speech detection approaches use time-frequency input representations such as STFT~\cite{baseline}, Mel-frequency cepstral coefficients \cite{mfcc_ref, schu2023using}, or Mel spectrograms\cite{mel_ref_vae}. These representations are then processed with architectures like convolutional neural networks (CNNs)\cite{baseline}, recurrent neural networks~\cite{lstm_patho_ref}, or autoencoders~\cite{mel_ref_vae} to learn pathology-discriminant cues and perform automatic pathological speech detection. Moreover, with the success of self-supervised foundation models like wav2vec 2.0~\cite{w2v2} for several downstream tasks~\cite{suberb_bench}, researchers have started leveraging embeddings obtained from these models and combining them with classifiers such as multi-layer perceptrons \cite{amiri2024adversarial} for automatic pathological speech detection\cite{ w2v2_different_layer}. Although wav2vec 2.0-based approaches generally outperform those using time-frequency input representations, the CNN-based approach operating on STFT  magnitude spectrograms has also demonstrated strong performance~\cite{amiri2024adversarial}.

\subsection{Multimodal LLMs}
Pretrained LLMs have demonstrated SOTA performance in downstream natural language understanding tasks~\cite{radford2019language,brown2020language}. 
More recently, LLMs have been extended to process multiple modalities, such as images and audio, alongside text prompts~\cite{gpt4,team2023gemini,alayrac2022flamingo,tangsalmonn,barrault2023seamlessm4t}, enabling zero-shot or few-shot learning. Among these multimodal models, GPT-4o~\cite{gpt4} (aka ChatGPT-4o) has achieved SOTA performance on various multimodal benchmarks, making it the model of choice for our study.

%Based on the significant performance of Transformer~\cite{vaswani2017attention} for various tasks in natural language processing, several work studied scaling of training data and network structure to train large language models (LLMs). The pretrained LLMs could surpass SOTA models in downstream tasks for natural language understanding~\cite{radford2019language,brown2020language}. Following remarkable performance  of LLMs, several models were proposed that can also take other modalities (e.g., image, audio, etc.) as input along with a text prompt~\cite{gpt4,team2023gemini,alayrac2022flamingo,tangsalmonn,barrault2023seamlessm4t}, and  be used for zero-shot or few-shot learning. Among multimodal LLMs, GPT-4o~\cite{gpt4} (aka ChatGPT-4o) has achieved SOTA performance on various multimodal benchmarks, and therefore we use this model in our study. 
% \textcolor{red}{Hatef: write a short servey}

\section{ Proposed ChatGPT-4o-based Approach}

Fig.~\ref{fig:measures} provides a schematic illustration of the proposed ChatGPT-4o-based approach for pathological speech detection. 
As depicted, the model is first provided with a system prompt describing the classification task, input representation, and number of reference samples. We then present the LLM with several reference samples (depending on the number of shots used) from both control and pathological speakers. Using this contextual knowledge, the model classifies a given test sample and generates an explanation for its decision. This added interpretability enhances trust in the classification process.

This approach enables a broad range of analyses, which we explore in an ablation study. In Section \ref{sec: exp}, we evaluate the performance by framing the task specified in the system prompt as either a general audio classification task (cf. Fig. \ref{fig:measures} and Section~\ref{sec: main}) or a dysarthria classification task (cf. Section~\ref{sec: task}). Additionally, we investigate the effect of requesting both a classification score between $0$ and $1$ ($0$: Controls, $1$: Patients with Cerebral Palsy) along with an explanation (cf. Fig. \ref{fig:measures} and Section~\ref{sec: main}) versus requesting only a classification score (cf. Section~\ref{sec: detail}). Finally, we analyze the influence of the input representation on performance by evaluating the ChatGPT-4o-audio-preview model with raw speech input (cf. Section~\ref{sec: raw}).

\begin{table*}[!ht]
\caption{Speaker-level accuracy of the proposed ChatGPT-4o-based model and the CNN-based baseline. The ChatGPT-4o model is evaluated under different few-shot scenarios. For a fair comparison, in addition to the baseline model trained on the entire training set, we also consider baseline models trained only on data from the same speakers included in the few-shot setup.
}
\label{tab: results}
\begin{center}
\begin{small}
\begin{sc}
\begin{tabularx}{\linewidth}{X|ccccccc}
\toprule

\multirow{1}{*}{Model} &\multicolumn{3}{|c}{ChatGPT-4o-based Network} & \multicolumn{4}{|c}{CNN-based baseline} \\

\multirow{1}{*}{Shot/Speaker number } & \multicolumn{1}{|c}{1 Shot} & \multicolumn{1}{|c}{3 Shot} &  \multicolumn{1}{|c}{5 Shot}& \multicolumn{1}{|c}{2 spk} & \multicolumn{1}{|c}{6 spk} & \multicolumn{1}{|c}{10 spk} & \multicolumn{1}{|c}{Full Data} \\

\midrule

\multirow{1}{*}{Accuracy (\%)} 
& \multicolumn{1}{|c}{$70.2 \pm 3.4$} & \multicolumn{1}{|c}{$82.1 \pm 0.0$} & \multicolumn{1}{|c}{$85.7 \pm 0.0$} & \multicolumn{1}{|c}{$84.5 \pm 1.6$} & \multicolumn{1}{|c}{$88.1 \pm 1.6$} & \multicolumn{1}{|c}{$90.5 \pm 1.6 $} &\multicolumn{1}{|c}{$95.2 \pm 1.6$} \\

\bottomrule
\end{tabularx}
\end{sc}
\end{small}
\end{center}
\vspace{-.2cm}
\end{table*} 

\section{Experimental Settings}
%In this section we describe the experimental settings used for our study.
\subsection{Database}
\label{sec: database}
In this paper, we use the Noise Reduced UA-Speech Dysarthria Dataset \cite{ua-denoise}\footnote{Under Attribution-NonCommercial 4.0 International (CC BY-NC 4.0) license.}, which is a denoised version of the UA-Speech~\cite{uaspeech} dataset. The dataset includes recordings from 16 speakers with Cerebral Palsy ($4$ females, $11$ males) and $13$ control speakers ($4$ females, $9$ males).
Each speaker utters various common words (CW), uncommon words (UW), commands (C), letters (L), and digits (D).
The recordings are acquired using a $7$-channel microphone array with a sampling frequency of $44.1$ kHz. Recordings are downsampled to 16 kHz. For the following experiments, we use recordings from the arbitrarily selected $5$-th channel.

Prior research \cite{schu2023using} has shown that the UA-Speech dataset exhibits significant differences between control and pathological recordings due to variations in recording setups and noise conditions. These differences are easier for DL-based approaches to learn than pathology-discriminant cues \cite{schu2023using}, yielding an unconventionally high accuracy of these approaches on this dataset.
Although we use the denoised version of the UA-Speech dataset, non-pathology-related differences persist, as enhancement methods introduce distortions that depend on noise conditions. As a result, the Noise Reduced UA-Speech dataset is also not optimal for automatic pathological speech detection experiments. However, due to licensing restrictions, it is, to our knowledge, the only dataset we can redistribute to third parties (i.e., upload to ChatGPT-4o). For this reason, we have chosen to use it in this paper despite its limitations.

\subsection{CNN Baseline Model}
As previously mentioned, this paper focuses on analyzing ChatGPT-4o's ability to process STFT magnitude spectrograms. For a fair comparison, we use a SOTA CNN model that also operates on STFT magnitude spectrograms as our baseline~\cite{baseline}.
Since the CNN model accepts only fixed-size inputs, we split each utterance into $500$ ms segments with an overlap of
$250$ ms. The STFT of these segments is computed using a $10$ ms Hanning window without overlap and the logarithm of the STFT magnitude is used as input representation. Input representations are normalized through a LayerNorm function ($\mu = 0, \sigma= 1$). The CNN architecture we use is adopted from~\cite{baseline}.

%Following~\cite{baseline}, we adopt a CNN with two convolutional layers ($64$ channels and kernel sizes of $2 \times 2$ and $3 \times 3$ in the first and second layer, respectively).
%Each convolutional layer is followed by batch normalization, a ReLU activation function, and max pooling with a $2 \times 2$ kernel. 
%Dropout ($p=0.5$) is applied after the second layer, and a linear layer (input size: $13376$, output size: $2$) is used for pathological speech detection.

For training and evaluation, we use a leave-one-speaker-out approach. 
In each fold, one speaker is used as a test speaker, while the remaining speakers are divided into training and validation sets with a $9:1$ ratio. The model is trained using the Adam optimizer with a learning rate of $0.001$ and a weight decay of $5 \times 10^{-3}$.

\subsection{ChatGPT-4o Setup}
To use ChatGPT-4o in an in-context learning scenario, we first define a system message that outlines the classification task, input representation, and number of reference samples per class. Next, we present the reference samples for each group based on the selected number of shots for in-context learning. Finally, we provide the test sample to the model. 

For selecting the reference samples, we strive to maintain a balanced distribution. We ensure that the number of speakers from each gender is equal in both classes and select identical utterances for each group. For instance, if a male speaker utters a specific common word in the control group, there will also be a male speaker uttering the same word in the pathological group. For the test samples, we randomly select two utterances from each of the five groups of recordings (i.e., CW, UW, C, D, and L), resulting in a total of $10$ test samples.
%To ensure comparability with the CNN-based results, we also adopt a leave-one-speaker-out approach for ChatGPT.
In this setup, each test speaker is evaluated separately, using reference samples from (a subset of) the remaining speakers.

\subsection{Performance Evaluation}
To evaluate the performance of the considered approaches, we consider the speaker-level accuracy. For the baseline model, we compute soft labels by passing the network's output through a softmax function for all segments belonging to each speaker. The speaker-level decision is then made through soft voting based on the scores of all these segments. For the ChatGPT-4o-based approach, the speaker-level decision is made through soft voting of the classification scores obtained for of all utterances belonging to the speaker. 

Since ChatGPT-4o is non-deterministic, we repeat each experiment three times on the same data and report the mean and standard deviation of the results. For the baseline model, we train all networks three times with different random seed initializations, and similarly report the mean and standard deviation of the accuracy.

\section{Experimental Results}
\label{sec: exp}
\subsection{Performance of Proposed and Baseline Models}
\label{sec: main}
In the following, the performance of the proposed ChatGPT-4o-based model is compared to the performance of the SOTA CNN-based baseline.
To this end, we perform in-context learning for the ChatGPT-4o-based model with different number of shots, i.e., $1$, $3$, and $5$.
We frame the task as a general audio classification task (cf. System Prompt in Fig.~\ref{fig:measures}) and request both a classification score and an
explanation of the decision for the test sample (cf. Test Prompt in Fig.~\ref{fig:measures}).
To ensure a fair comparison to the baseline CNN model, we consider training different CNN models for each shot by using only data from the reference speakers selected for the ChatGPT-4o-based model.\footnote{Please note that these CNN models use all the data available from the reference speakers, not just the reference samples used for the ChatGPT-4o-based model. This is necessary to provide sufficient training data for the CNN model.}
For completeness, we also consider the results obtained when all the training data is used for the CNN model.

Table \ref{tab: results} presents the obtained results.
It can be observed that as expected, both models show a performance improvement as they are exposed to more training data. More importantly, the ChatGPT-4o-based model demonstrates promising results compared to the SOTA baseline. 
While the ChatGPT-4o-based model lags behind the CNN model in terms of accuracy, its ability to provide explanations for its decisions (cf. exemplary Prediction in Fig.~\ref{fig:measures}) enhances interpretability, making it more suitable for critical applications.
In the following subsections, we analyze the impact of different settings on the performance of the ChatGPT-4o-based model.

\subsection{Impact of System Prompt}
\label{sec: task}
The system prompt defines the model's behavior by specifying the task and instructing it to classify test samples based on reference samples. 
As previously mentioned, there are two primary ways to define the task in the system prompt: (i) as an audio classification task without further details as in Section \ref{sec: main} or (ii) as a dysarthria classification task, where the model is provided with patient characteristics and asked to classify based on both the reference samples and the given description. In describing dysarthria, we highlight key symptoms such as articulation deficiencies, vowel distortions, reduced loudness variation, hypernasality, and syllabification issues~\cite{duffy2012motor}. To examine the impact of the system prompt on performance, we repeat the ChatGPT-4o-based experiments from Section~\ref{sec: main} using a system prompt that includes a description of dysarthria. 

Table \ref{tab: ablation} presents the obtained results using a system prompt that describes dysarthria (Dysarthria-specific prompt). For ease of comparison, it also includes the ChatGPT-4o-based results from Section~\ref{sec: main}, where the system prompt frames the task as general audio classification.
The results show that using a system prompt specifically describing dysarthria leads to a performance degradation, regardless of the number of shots considered.
As described in Section \ref{sec: database}, we expect the denoised UA-Speech database to  contain pathology-unrelated differences between the two groups of speakers that are considerably easier to learn than pathology-discriminant cues.
We believe that when a description of the pathology is included in the system prompt, the ChatGPT-based model shifts its focus to these characteristics, rather than relying on spurious pathology-unrelated cues. 
Therefore, we hypothesize that this degradation in performance is due to the model focusing on genuine pathological cues rather than unintended artifacts. Further investigation is needed to fully understand the network's decision-making process in this context.

% \begin{table}[h!]
%   \caption{The accuracy of the ChatGPT-4o-based approach when providing a description of dysarthria in the system prompt.}
%   \label{tab: task_sys}
%   \centering
%       \begin{tabularx}{\linewidth}{X|ccc}
%     \toprule
%     \textbf{Shot} & 1-Shot& 3-Shot& 5-Shot \\
%     \midrule
%     \textbf{Accuracy (\%)} &  $60.7 \pm 2.9$ & $69.0 \pm 1.6$ & $76.2 \pm 1.6$
%     \\
%     \bottomrule
%   \end{tabularx}
% \end{table}

%We hypothesize that since these samples were chosen randomly, they might not align well with the description provided to the network. We believe that by carefully selecting the samples and explicitly mentioning the characteristics present in each, we can better guide the model and leverage the prior knowledge of LLMs.

\begin{table}[h!]
  \caption{Impact of different settings on the performance of the proposed approach under different few-shot scenarios.}
  \label{tab: ablation}
  \centering
      \begin{tabularx}{\linewidth}{X|ccc}
    \toprule
    \textbf{Accuracy (\%)} & 1-Shot& 3-Shot& 5-Shot \\
    \midrule
    Setting from Section  \ref{sec: main} & $70.2 \pm 3.4$ &$82.1 \pm 0.0$& $85.7 \pm 0.0$
    \\
    \scriptsize	{Dysarthria-specific prompt} & $60.7 \pm 2.9$ & $69.0 \pm 1.6$ & $76.2 \pm 1.6$
    \\
    \scriptsize	{Non-detailed response} &  $64.3 \pm 0.0$ &$75.0 \pm 0.0$& $79.8 \pm 3.4$
    \\
    \scriptsize	{Raw speech input} & $52.4\pm 6.7$&$60.7 \pm 5.8$& $67.9 \pm 2.9$
    \\
    \bottomrule
  \end{tabularx}
\end{table}

\subsection{Impact of Non-detailed Response}
\label{sec: detail}
At test time, we prompt the LLM to classify the test STFT representation into one of the predefined classes. This can be done in two ways: (i) requesting only a classification score, or (ii) asking the model to provide both a score and an explanation for its decision as in Section \ref{sec: main}. 
To examine the impact of this choice on performance, we repeat the ChatGPT-4o-based experiments from Section~\ref{sec: main} requesting only a classification score.
The results for this setting are reported in Table \ref{tab: ablation} (Non-detailed response).
We observe a degradation in the performance of the proposed method when it is asked to return only a classification score, compared to when it also provides an explanation for its prediction. This is because the model performs better when it follows a step-by-step reasoning process to generate its response. Similar findings have been reported in previous studies, where the chain-of-thought prompting technique has been shown to improve the performance of LLMs~\cite{wei2022chain}.

\subsection{Impact of Raw Speech Input}
\label{sec: raw}
To examine the impact of different input representations on performance, we repeat the experiments from Section \ref{sec: main} using raw speech input and the ChatGPT-4o-Audio-Preview model. The results are presented in Table \ref{tab: ablation} (Raw speech input).
As observed, there is a degradation in performance compared to when the STFT representation is used as input. While raw speech inherently contains more information than STFT, we conclude that ChatGPT-4o's vision capabilities are more advanced than its audio processing abilities, resulting in a better performance with STFT input.

% \subsection{Results Variance}

% Since ChatGPT-4o is not a deterministic model, it can generate different outputs for the same input across different runs. To examine the impact of this probabilistic effect on the final results, we run our proposed method multiple times and compute the standard deviation across three different runs for each shot scenario. These results are presented in Table \ref{tab: results}.

% As observed, while there is a small variation in the 1-shot scenario, no variation is present in the accuracy of the 3-shot and 5-shot scenarios. It is important to note that we do observe minor variations in utterance-level accuracy; however, these differences are not significant enough to impact speaker-level accuracy. We conclude that providing more samples to the ChatGPT-4o model increases its confidence in classifying test samples, which explains the absence of variation in higher-shot scenarios.

\section{Conclusion}

In this study, we explored the in-context learning capabilities of ChatGPT-4o for automatic pathological speech detection. Evaluating various shot settings on the denoised UA-Speech dataset, we found that our method delivers promising performance while also providing explanations for its decisions. These results highlight the potential of LLMs in this task and emphasize the need for further research to fully harness their capabilities. In future work, we will investigate the quality of the explanations provided by GPT-4o and explore approaches to enhance their effectiveness.

% In this study, we explored the in-context learning capabilities of ChatGPT-4o for automatic pathological speech detection. We evaluated our approach across different shot settings and compared it to the SOTA method on the UA-Speech dataset. Our results demonstrate that the proposed method achieves promising and competitive performance while also providing explainability by detailing the reasoning behind its decisions.

% To further analyze our method, we conducted an ablation study. We observed that while ChatGPT-4o is inherently non-deterministic, its performance stabilizes when exposed to more samples, resulting in consistent outputs in higher-shot scenarios. Additionally, we found that explicitly describing the task in the system prompt led to a slight performance degradation, likely because the model focused more on actual pathological cues rather than spurious ones. Finally, our results indicate that requesting an explanation for the model’s classification leads to improved accuracy. This aligns with previous findings, where chain-of-thought prompting has been shown to enhance the performance of LLMs.

% Our promising results suggest that further research should explore this direction to fully leverage the potential of LLMs for pathological speech detection.

\vspace{12pt}

\bibliographystyle{IEEEtran}
\bibliography{mybib}

% Generated by IEEEtran.bst, version: 1.14 (2015/08/26)
\begin{thebibliography}{10}
\providecommand{\url}[1]{#1}
\csname url@samestyle\endcsname
\providecommand{\newblock}{\relax}
\providecommand{\bibinfo}[2]{#2}
\providecommand{\BIBentrySTDinterwordspacing}{\spaceskip=0pt\relax}
\providecommand{\BIBentryALTinterwordstretchfactor}{4}
\providecommand{\BIBentryALTinterwordspacing}{\spaceskip=\fontdimen2\font plus
\BIBentryALTinterwordstretchfactor\fontdimen3\font minus
  \fontdimen4\font\relax}
\providecommand{\BIBforeignlanguage}[2]{{%
\expandafter\ifx\csname l@#1\endcsname\relax
\typeout{** WARNING: IEEEtran.bst: No hyphenation pattern has been}%
\typeout{** loaded for the language `#1'. Using the pattern for}%
\typeout{** the default language instead.}%
\else
\language=\csname l@#1\endcsname
\fi
#2}}
\providecommand{\BIBdecl}{\relax}
\BIBdecl

\bibitem{duffy2012motor}
J.~R. Duffy \emph{et~al.}, \emph{Motor speech disorders: Substrates,
  differential diagnosis, and management}.\hskip 1em plus 0.5em minus
  0.4em\relax Elsevier Health Sciences, 2012.

\bibitem{ziegler2012apraxia}
W.~Ziegler, I.~Aichert, and A.~Staiger, ``Apraxia of speech: Concepts and
  controversies,'' \emph{Journal of speech, language, and hearing research},
  vol.~55, no.~5, pp. S1485--S1501, 2012.

\bibitem{handcrafted_1}
I.~Kodrasi, M.~Pernon, M.~Laganaro, and H.~Bourlard, ``Automatic discrimination
  of apraxia of speech and dysarthria using a minimalistic set of handcrafted
  features.'' in \emph{Proc. Annual Conference of the International Speech
  Communication Association}, Oct. 2020, pp. 4991--4995.

\bibitem{handcrafted_2}
I.~Kodrasi and H.~Bourlard, ``Spectro-temporal sparsity characterization for
  dysarthric speech detection,'' \emph{IEEE/ACM Transactions on Audio, Speech,
  and Language Processing}, vol.~28, no.~6, pp. 1210--1222, June 2020.

\bibitem{handcrafted_3}
P.~Janbakhshi, I.~Kodrasi, and H.~Bourlard, ``Subspace-based learning for
  automatic dysarthric speech detection,'' \emph{IEEE Signal Processing
  Letters}, vol.~28, pp. 96--100, Jan. 2020.

\bibitem{transformer_ref}
A.~Vaswani, N.~Shazeer, N.~Parmar, J.~Uszkoreit, L.~Jones, A.~N. Gomez,
  L.~Kaiser, and I.~Polosukhin, ``Attention is all you need,'' in \emph{Proc.
  Advances in Neural Information Processing Systems}, California, USA, Dec.
  2017, pp. 6000--6010.

\bibitem{speech_ref}
A.~Graves, A.~Mohamed, and G.~Hinton, ``Speech recognition with deep recurrent
  neural networks,'' in \emph{Proc. IEEE International Conference on Acoustics,
  Speech and Signal Processing}, Vancouver, Canada, May 2013, pp. 6645--6649.

\bibitem{baseline}
P.~Janbakhshi and I.~Kodrasi, ``Experimental investigation on {STFT} phase
  representations for deep learning-based dysarthric speech detection,'' in
  \emph{Proc. IEEE International Conference on Acoustics, Speech and Signal
  Processing}, Philadelphia, USA, May 2022, pp. 6477--6481.

\bibitem{deep_learning_based}
------, ``Supervised speech representation learning for {P}arkinson’s disease
  classification,'' in \emph{Proc. ITG Conference on Speech Communication},
  Kiel, Germany, Sept. 2021, pp. 154--158.

\bibitem{mel_ref_vae}
------, ``Adversarial-free speaker identity-invariant representation learning
  for automatic dysarthric speech classification,'' in \emph{Proc. Annual
  Conference of the International Speech Communication Association}, Incheon,
  Korea, Sept. 2022, pp. 2138--2142.

\bibitem{mfcc_ref}
K.~L. Kadi, S.~A. Selouani, B.~Boudraa, and M.~Boudraa, ``Fully automated
  speaker identification and intelligibility assessment in dysarthria disease
  using auditory knowledge,'' \emph{Biocybernetics and Biomedical Engineering},
  vol.~36, no.~1, pp. 233--247, Jan. 2016.

\bibitem{schu2023using}
G.~Schu, P.~Janbakhshi, and I.~Kodrasi, ``On using the ua-speech and torgo
  databases to validate automatic dysarthric speech classification
  approaches,'' in \emph{ICASSP 2023-2023 IEEE International Conference on
  Acoustics, Speech and Signal Processing (ICASSP)}.\hskip 1em plus 0.5em minus
  0.4em\relax IEEE, 2023, pp. 1--5.

\bibitem{lstm_patho_ref}
J.~Millet and N.~Zeghidour, ``Learning to detect dysarthria from raw speech,''
  in \emph{Proc. IEEE International Conference on Acoustics, Speech and Signal
  Processing}, Brighton, UK, May 2019, pp. 5831--5835.

\bibitem{amiri2024test}
M.~Amiri and I.~Kodrasi, ``Test-time adaptation for automatic pathological
  speech detection in noisy environments,'' in \emph{2024 32nd European Signal
  Processing Conference (EUSIPCO)}.\hskip 1em plus 0.5em minus 0.4em\relax
  IEEE, 2024, pp. 86--90.

\bibitem{amiri_suppress}
------, ``Suppressing noise disparity in training data for automatic
  pathological speech detection,'' in \emph{2024 18th International Workshop on
  Acoustic Signal Enhancement (IWAENC)}, 2024, pp. 110--114.

\bibitem{interpret}
L.~Xu, J.~Liss, and V.~Berisha, ``Dysarthria detection based on a deep learning
  model with a clinically-interpretable layer,'' \emph{JASA Express Letters},
  vol.~3, no.~1, 2023.

\bibitem{latif2023sparkslargeaudiomodels}
\BIBentryALTinterwordspacing
S.~Latif, M.~Shoukat, F.~Shamshad, M.~Usama, Y.~Ren, H.~Cuayáhuitl, W.~Wang,
  X.~Zhang, R.~Togneri, E.~Cambria, and B.~W. Schuller, ``Sparks of large audio
  models: A survey and outlook,'' 2023. [Online]. Available:
  \url{https://arxiv.org/abs/2308.12792}
\BIBentrySTDinterwordspacing

\bibitem{awais2025foundation}
M.~Awais, M.~Naseer, S.~Khan, R.~M. Anwer, H.~Cholakkal, M.~Shah, M.-H. Yang,
  and F.~S. Khan, ``Foundation models defining a new era in vision: a survey
  and outlook,'' \emph{IEEE Transactions on Pattern Analysis and Machine
  Intelligence}, 2025.

\bibitem{fmbiometrics2025survey}
H.~O. Shahreza and S.~Marcel, ``Foundation models and biometrics: A survey and
  outlook,'' \emph{Authorea Preprints}, 2025.

\bibitem{gpt4}
J.~Achiam, S.~Adler, S.~Agarwal, L.~Ahmad, I.~Akkaya, F.~L. Aleman, D.~Almeida,
  J.~Altenschmidt, S.~Altman, S.~Anadkat \emph{et~al.}, ``Gpt-4 technical
  report,'' \emph{arXiv preprint arXiv:2303.08774}, 2023.

\bibitem{hassanpour2024chatgpt}
A.~Hassanpour, Y.~Kowsari, H.~O. Shahreza, B.~Yang, and S.~Marcel, ``Chatgpt
  and biometrics: an assessment of face recognition, gender detection, and age
  estimation capabilities,'' in \emph{2024 IEEE International Conference on
  Image Processing (ICIP)}.\hskip 1em plus 0.5em minus 0.4em\relax IEEE, 2024,
  pp. 3224--3229.

\bibitem{koubaa2023exploring}
A.~Koubaa, W.~Boulila, L.~Ghouti, A.~Alzahem, and S.~Latif, ``Exploring chatgpt
  capabilities and limitations: a survey,'' \emph{IEEE Access}, vol.~11, pp.
  118\,698--118\,721, 2023.

\bibitem{he2023can}
M.~He and P.~N. Garner, ``Can chatgpt detect intent? evaluating large language
  models for spoken language understanding,'' \emph{arXiv preprint
  arXiv:2305.13512}, 2023.

\bibitem{komaty2025exploring}
A.~Komaty, H.~O. Shahreza, A.~George, and S.~Marcel, ``Exploring chatgpt for
  face presentation attack detection in zero and few-shot in-context
  learning,'' \emph{Proceedings of the IEEE/CVF Winter Conference on
  Applications of Computer Vision}, 2025.

\bibitem{brown2020language}
T.~Brown, B.~Mann, N.~Ryder, M.~Subbiah, J.~D. Kaplan, P.~Dhariwal,
  A.~Neelakantan, P.~Shyam, G.~Sastry, A.~Askell \emph{et~al.}, ``Language
  models are few-shot learners,'' \emph{Advances in neural information
  processing systems}, vol.~33, pp. 1877--1901, 2020.

\bibitem{ua-denoise}
``Noise reduced uaspeech dysarthria dataset,''
  https://www.kaggle.com/datasets/aryashah2k/noise-reduced-uaspeech-dysarthria-dataset/.

\bibitem{amiri2024adversarial}
M.~Amiri and I.~Kodrasi, ``Adversarial robustness analysis in automatic
  pathological speech detection approaches,'' in \emph{Proc. Annual Conference
  of the International Speech Communication, Rhodes Islands, Greece}, 2024, pp.
  1415--1419.

\bibitem{w2v2}
A.~Baevski, Y.~Zhou, A.~Mohamed, and M.~Auli, ``wav2vec 2.0: A framework for
  self-supervised learning of speech representations,'' in \emph{Proc. Annual
  Conference on Neural Information Processing Systems}, Virtual, Dec. 2020, pp.
  12\,449--12\,460.

\bibitem{suberb_bench}
S.~wen Yang, P.-H. Chi, Y.-S. Chuang, C.-I.~J. Lai, K.~Lakhotia, Y.~Y. Lin,
  A.~T. Liu, J.~Shi, X.~Chang, G.-T. Lin, T.-H. Huang, W.-C. Tseng, K.~tik Lee,
  D.-R. Liu, Z.~Huang, S.~Dong, S.-W. Li, S.~Watanabe, A.~Mohamed, and
  H.~yi~Lee, ``{SUPERB: Speech Processing Universal PERformance Benchmark},''
  in \emph{Proc. Interspeech 2021}, 2021, pp. 1194--1198.

\bibitem{w2v2_different_layer}
F.~Javanmardi, S.~Tirronen, M.~Kodali, S.~R. Kadiri, and P.~Alku,
  ``Wav2vec-based detection and severity level classification of dysarthria
  from speech,'' in \emph{ICASSP 2023-2023 IEEE International Conference on
  Acoustics, Speech and Signal Processing (ICASSP)}.\hskip 1em plus 0.5em minus
  0.4em\relax IEEE, 2023, pp. 1--5.

\bibitem{radford2019language}
A.~Radford, J.~Wu, R.~Child, D.~Luan, D.~Amodei, and I.~Sutskever, ``Language
  models are unsupervised multitask learners,'' \emph{OpenAI blog}, 2019.

\bibitem{team2023gemini}
G.~Team, R.~Anil, S.~Borgeaud, J.-B. Alayrac, J.~Yu, R.~Soricut, J.~Schalkwyk,
  A.~M. Dai, A.~Hauth, K.~Millican \emph{et~al.}, ``Gemini: a family of highly
  capable multimodal models,'' \emph{arXiv preprint arXiv:2312.11805}, 2023.

\bibitem{alayrac2022flamingo}
J.-B. Alayrac, J.~Donahue, P.~Luc, A.~Miech, I.~Barr, Y.~Hasson, K.~Lenc,
  A.~Mensch, K.~Millican, M.~Reynolds \emph{et~al.}, ``Flamingo: a visual
  language model for few-shot learning,'' \emph{Advances in neural information
  processing systems}, vol.~35, pp. 23\,716--23\,736, 2022.

\bibitem{tangsalmonn}
C.~Tang, W.~Yu, G.~Sun, X.~Chen, T.~Tan, W.~Li, L.~Lu, M.~Zejun, and C.~Zhang,
  ``Salmonn: Towards generic hearing abilities for large language models,'' in
  \emph{The Twelfth International Conference on Learning Representations},
  2024.

\bibitem{barrault2023seamlessm4t}
L.~Barrault, Y.-A. Chung, M.~C. Meglioli, D.~Dale, N.~Dong, P.-A. Duquenne,
  H.~Elsahar, H.~Gong, K.~Heffernan, J.~Hoffman \emph{et~al.},
  ``Seamlessm4t-massively multilingual \& multimodal machine translation,''
  \emph{arXiv preprint arXiv:2308.11596}, 2023.

\bibitem{uaspeech}
H.~Kim, M.~Hasegawa-Johnson, A.~Perlman, J.~Gunderson, T.~S. Huang, K.~Watkin,
  and S.~Frame, ``Dysarthric speech database for universal access research,''
  in \emph{Interspeech 2008}, 2008, pp. 1741--1744.

\bibitem{wei2022chain}
J.~Wei, X.~Wang, D.~Schuurmans, M.~Bosma, F.~Xia, E.~Chi, Q.~V. Le, D.~Zhou
  \emph{et~al.}, ``Chain-of-thought prompting elicits reasoning in large
  language models,'' \emph{Advances in neural information processing systems},
  vol.~35, pp. 24\,824--24\,837, 2022.

\end{thebibliography}

\end{document}